\documentclass[aps,prl,twocolumn,superscriptaddress,nofootinbib,superscriptaddress]{revtex4-2}
 
\pdfoutput=1

\usepackage{slashed}
\usepackage{colortbl}
\usepackage{latexsym}
\usepackage{amssymb}
\usepackage{amsmath}
\usepackage{graphicx}
\usepackage{xcolor}
\usepackage{adjustbox}
\usepackage{booktabs}
\usepackage{multirow}
\usepackage{rotating}
\usepackage{hyperref}
\usepackage{mathtools}
\usepackage{calrsfs}
\usepackage{stackrel}
\usepackage{xspace}
\usepackage{bm}
\usepackage[capitalise]{cleveref}
\usepackage{ulem}

\newcommand{\msbar}{\ensuremath{
\overline{\text{MS}}\xspace}}
\newcommand{\GeV}{\ensuremath{\text{GeV}\xspace}}

\begin{document}
\begin{flushright}
CERN-TH-2025-220
\end{flushright}
\title{Running Couplings in High-Temperature Effective Field Theory}

\author{Mikael Chala}
\email{mikael.chala@ugr.es}
\affiliation{Departamento de F\'isica Te\'orica y del Cosmos, Universidad de Granada, E--18071 Granada, Spain}
\author{Andrii Dashko}
\email{andrii.dashko@ugr.es}
\affiliation{Departamento de F\'isica Te\'orica y del Cosmos, Universidad de Granada, E--18071 Granada, Spain}
\author{Guilherme Guedes}
\email{guilherme.guedes@cern.ch}
\affiliation{CERN, Theoretical Physics Department, Esplanade des Particules 1, Geneva 1211, Switzerland}

\begin{abstract} 
In this work, we study the renormalization-group evolution of parameters in the three-dimensional effective field theory (3D EFT) that describes the thermally driven electroweak phase transition of the Higgs field. 
We consider tree-level and radiatively generated barriers induced by beyond the Standard Model physics, enabling a first-order phase transition at and below the soft scale, respectively.
For each case, we compute the two-loop running of the 3D EFT couplings, including the effects of the leading nonrenormalizable terms.
We then analyze how the new contributions to the beta functions compare with those in the super-renormalizable case, highlighting their impact on perturbative computations of the scalar potential, which describes the vacuum structure of the theory.
By incorporating higher-order corrections in the mass parameter evolution, as well as the running of other effective operators, we set the stage for testing their impact on phase transition dynamics in lattice simulations.
\end{abstract}

\maketitle

\newpage

\section{Introduction} 
Phase transitions (PTs) are ubiquitous in nature, from the condensation of water vapor to the magnetization of materials. Strikingly, however, the Standard Model (SM) itself does not feature any cosmological PT of first order~\cite{Kajantie:1996mn,Aoki:2006we}. The discovery of such a transition would therefore be a major breakthrough, pointing unambiguously to physics beyond the SM (BSM). This possibility is particularly timely, as present and future gravitational wave (GW) detectors may be sensitive to the stochastic backgrounds generated by cosmological PTs~\cite{Caprini:2019egz,Caprini:2024hue}. Motivated by this, significant effort is being devoted to refining the theoretical predictions of relevant PT parameters, particularly for those taking place at the electroweak (EW) scale~\cite{Gould:2019qek,Hirvonen:2021zej,Lofgren:2021ogg,Ekstedt:2022zro,Schicho:2022wty,Laurent:2022jrs,Gould:2023ovu,Ekstedt:2023sqc,Athron:2023xlk,Dashko:2024anp,Ekstedt:2024etx,Ekstedt:2024fyq,Tian:2024ysd,Niemi:2024vzw,Chala:2024xll,Branchina:2025adj,Ai:2025bjw,Chen:2025ksr,Branchina:2025jou,Bernardo:2025vkz,Carena:2025flp,Chala:2025aiz,Chala:2025oul,Navarrete:2025yxy,Croon:2020cgk,Gould:2021oba,Kierkla:2025qyz,Li:2025kyo}.

The state-of-the-art framework to study thermally driven PTs is dimensional reduction (DR)~\cite{Ginsparg:1980ef,Farakos:1994kx,Kajantie:1995dw}, whereby the nonzero Matsubara modes are integrated out to obtain a purely three-dimensional (3D) effective field theory (EFT) without fermions.
Automated tools have been developed to make the use of DR more accessible for phenomenological studies of PTs in BSM scenarios~\cite{Ekstedt:2022bff}.
This 3D EFT, even when restricted to SM fields alone, is highly predictive: not only does it faithfully capture the infrared (IR) dynamics of the SM at high temperatures ($T>m$, where $m$ stands for the light mass of the nucleating field and $T$ is the temperature), but it also serves as a universal description of a broad class of ultraviolet (UV) extensions. In the leading-order (LO) truncation of the $m/T$ expansion (considering only super-renormalizable scalar interactions), the EW sector of this EFT has been simulated on the lattice~\cite{Kajantie:1995kf,Kajantie:1996mn,Gould:2019qek}, with the conclusion that the SM cannot realize a PT strong enough to source observable GWs. 
More recently, perturbative analyses have been shown to reproduce well the PT before the critical point~\cite{Ekstedt:2024etx}, which can be only seen in lattice.

The situation changes once higher-order terms in $m/T$ expansion are retained. These manifest as higher-dimensional operators in the 3D EFT, and recent studies~\cite{Chala:2024xll,Bernardo:2025vkz,Chala:2025aiz,Chala:2025oul,Biekotter:2025npc} have demonstrated that their impact is essential to capture strong PTs~\footnote{There are also promising approaches~\cite{Navarrete:2025yxy,Bittar:2025lcr} aiming to avoid the $m/T$ expansion altogether, though they remain at an early stage.}. Precise knowledge of the renormalization group (RG) evolution of the 3D EFT coefficients is fundamental both in perturbative calculations (addressing renormalization-scale dependence from which other 4D methods suffer~\cite{Croon:2020cgk,Biekotter:2025npc}) and in the study of the 3D theory in the lattice~\cite{Farakos:1994kx,Kajantie:1995kf}. The computation of such RG equations constitutes the main objective of the present work. Namely, these results pave the way for the future inclusion of effective parameters (in particular, six-Higgs interactions) in lattice studies of PTs. Our results are also of direct use in DR computations, as IR divergences arising from the matching procedure must be consistently reproduced by the UV divergences we compute within the EFT. Furthermore, these results are also relevant for the study of critical phenomena through $\epsilon$ expansion~\cite{ODwyer:2007brp,Kapoor:2021lrr,Henriksson:2025kws}.

The paper is organized as follows.
In~\cref{sec:dimredtheory} we introduce our notation as well as the assumed power counting for 3D parameters.
We summarize the technical aspects of the computation of the relevant counterterms and beta functions for the effective couplings in~\cref{sec:computation}. In~\cref{sec:scalarpotential} we examine the impact of the running on the scalar potential, where we also compare with the super-renormalizable case. We conclude in~\cref{sec:conclusions}.
The case of the radiatively generated barrier is discussed in \cref{app:rgb}.
The contributions from the $U(1)_Y$ gauge sector are presented in \cref{app:g1}.

\section{Dimensionally reduced theory}
\label{sec:dimredtheory}
We are interested in the dynamics of the nucleating field that takes place below the hard scale, characterized by the 
temperature $T$,
and where only SM fields matter.
It is described by a dimensionally reduced theory, where temperature is no longer a dynamical parameter and enters the theory only through Wilson coefficients~\cite{Farakos:1994kx,Kajantie:1995dw}.
Considering, for simplicity, only the $SU(2)$ part of the SM for now (the contributions from the $U(1)_Y$ coupling $g'_3$ can be found in~\cref{app:g1}), the 3D EFT Lagrangian can be written as,
\begin{widetext}
\begin{equation}
    \begin{split}
\label{eq:lagrangian3d}
    \mathcal{L} &= \frac{1}{4} W_{rs}^I W_{rs}^I + (D_r\phi)^\dagger (D_r\phi) + m_3^2 |\phi|^2 + \lambda_3|\phi|^4 + c_{\phi^6} |\phi|^6\\
    &+ c_{\phi^4D^2}^{(1)} |\phi|^2\square|\phi|^2 + c_{\phi^4 D^2}^{(2)} (\phi^\dagger D_r\phi)^\dagger \phi^\dagger D_r\phi  + c_{\phi^8}|\phi|^8 + c_{\phi^2 W^2}|\phi|^2 W^I_{rs}W^I_{rs}\textcolor{gray}{+ r_{\phi^4 D^2}^{(3)} |\phi|^2 (D_r\phi)^\dagger D_r \phi}\,\\
    &+ \frac{1}{2} (D_rW_{0}^I) (D_rW_{0}^I) + \frac{1}{2}m_D^2 W_{0}^I W_{0}^I + \lambda_{W_0^4} (W_{0}^I W_{0}^I)^2+ \lambda_{\phi^2W_0^2} |\phi|^2W_{0}^I W_{0}^I \\
    &+ c_{W_0^6} (W_{0}^I W_{0}^I)^3+c_{\phi^2 W_0^4} |\phi|^2(W_{0}^I W_{0}^I)^2+c^{(1)}_{\phi^4 W_0^2} |\phi|^4(W_{0}^I W_{0}^I)+c^{(2)}_{\phi^4 W_0^2}(\phi^\dagger \sigma^I \phi)(\phi^\dagger \sigma^J\phi)(W_{0}^I W_{0}^J)
    \\
    & + c^{(1)}_{\phi^6 W_0^2} |\phi|^6(W_{0}^I W_{0}^I) + c^{(2)}_{\phi^6 W_0^2}|\phi|^2(\phi^\dagger \sigma^I \phi)(\phi^\dagger \sigma^J\phi)(W_{0}^I W_{0}^J) 
    \\
    &+ c^{(1)}_{\phi^2 W_0^2 D^2} |\phi|^2(D_r W_{0}^I) (D_r W_{0}^I)+ c^{(2)}_{\phi^2 W_0^2 D^2} \epsilon^{IJK} (\phi^\dagger \sigma^I i \overleftrightarrow{D}_{r} \phi)(W_{0}^J D_r W_{0}^K) + \textcolor{gray}{ r^{(3)}_{\phi^2 W^2_0 D^2} \phi^\dagger ( -i \overleftrightarrow{D}_r) \phi (D_r W_0^I) W_0^I} 
    \\
    &+\textcolor{gray}{ r^{(4)}_{\phi^2 W^2_0 D^2}  \left[ D^2(\phi^\dagger)\phi + \phi^\dagger D^2\phi \right] W_0^I W_0^I } + \textcolor{gray}{ r^{(5)}_{\phi^2 W^2_0 D^2} D^2 (W_0^I) W_0^I |\phi|^2} + \textcolor{gray}{r^{(6)}_{\phi^2 W^2_0 D^2} \epsilon^{IJK}D^2 (W_0^I) W_0^J \phi^\dagger \sigma^K \phi} +\cdots\, ,
    \end{split}
\end{equation}
\end{widetext}
where the ellipses stand for terms with higher-dimensional operators.
$\phi$ represents the Higgs doublet,  
$W$ denotes the $SU(2)_L$ gauge boson with coupling $g_3$ and $W_0$ is its temporal component.
Note that among dimension-four operators, we only consider those containing fewer than four gauge fields, since operators involving 
more gauge fields are 
suppressed in the DR procedure by extra powers of the gauge coupling.
In addition, we omit formally present operators of fractional dimensions, e.g., $|\phi|^2 (\phi^\dagger \sigma^I\phi)W_{0}^I$, as they are not generated in DR. 

Operators in \textcolor{gray}{gray} in \cref{eq:lagrangian3d} are redundant, which can be removed with a proper field redefinition~\cite{Chala:2025aiz}, discussed below.

If the nucleation of the
Higgs field occurs below the soft scale, characterized by the screening mass $m_{\text{D}}\sim gT$, 
where $g$ is a gauge coupling, then any degrees of freedom at the soft scale, including electric gauge fields, can be further integrated out.
The 3D EFT, valid at scales lower than the soft scale, has the same form as in \cref{eq:lagrangian3d}, but without terms including temporal gauge bosons (see \cref{app:rgb}). 

For completeness, we recall that within the 3D EFT, the energy dimensions of the fields and couplings are $[\phi]=[W_{\mu}]=1/2$ and $[m_3^2]=[m_D^2]=2$, $[\lambda_3]=[\lambda_{\phi^2 W_0^2}]=[\lambda_{W_0^4}]=1$, $[g_3]=1/2$, $[c_{\phi^6}]=[c_{\phi^4 W_0^2}]=[c_{\phi^2 W_0^4}]=[c_{W_0^6}]=0$, $[c_{\phi^8}]=[c_{\phi^4D^2}] = [c_{\phi^2 W^2}] = [c_{\phi^6 W_0^2}] = [c_{\phi^2 W_0^2 D^2}] = -1$.

Different hierarchies among the effective couplings in \cref{eq:lagrangian3d} lead to different PT patterns. 
These can include both tree-level barriers~\cite{Grojean:2004xa,Ham:2004zs} as well as radiatively generated barriers~\cite{Arnold:1992rz,Ekstedt:2022zro}. Possible realizations of the first-order PT within the SM EFT were discussed in Ref.~\cite{Camargo-Molina:2024sde}.
When discussing the phenomenological impact of higher-order running, 
we focus on the case of the tree-level barrier in the main text,
while relegating
the case of the radiatively generated barriers to
\cref{app:rgb}. In both scenarios,
we further assume that the theory in \cref{eq:lagrangian3d} is a result of the DR of the SMEFT~\cite{Buchmuller:1985jz, Isidori:2023pyp}. The latter is itself an effective description of generic BSM physics, characterized by a UV gauge coupling $\tilde{g}$ and valid up to scales $\Lambda \gtrsim T$.

The SMEFT couplings, denoted with a bar, are subsequently matched to the effective couplings in \cref{eq:lagrangian3d}~\cite{Chala:2025aiz}. Parametrically, the leading matching relations
for the Higgs-only operators 
are,~\footnote{We assume the following power counting for nonrenormalizable 4D terms~\cite{Isidori:2023pyp}: $\bar{c}_{\phi^6} \sim \tilde{g}^4/\Lambda^2, \bar{c}_{\phi^4D^2} \sim \tilde{g}^2/\Lambda^2,  \bar{c}_{\phi^2 W^2} \sim \tilde{g}^2 g^2/\Lambda^2, \bar{c}_{\phi^8} \sim \tilde{g}^6/\Lambda^4$. One can also consider an additional $(1/16\pi^2)$ suppression factor for $\bar{c}_{\phi^2 W^2}$ under the assumption of a weakly coupled completion of the SMEFT, as this coefficient is only generated at one loop~\cite{Einhorn:2013kja}} 
\begin{equation}
\label{eq:all-scalings-soft}
        \begin{split}
            m_3^2&\sim \bar{m}^2+g^2T^2,  \\
            \lambda_3 &\sim \bar{\lambda} T + \bar{c}_{\phi^6} T^3, \\
            g_3^2 &\sim g^2 T,\\
            c_{\phi^6} &\sim T^2\bar{c}_{\phi^6} \sim \tilde{g}^4 \frac{T^2}{\Lambda^2},\\
            c_{\phi^8} &\sim T^3 \bar{c}_{\phi^8} \sim  \tilde{g}^6 \frac{T^3}{\Lambda^4}, \\ c_{\phi^4D^2} &\sim  T \bar{c}_{\phi^4D^2} \sim  \tilde{g}^2 \frac{T}{\Lambda^2}, \\
        \end{split}
\end{equation}

The presence of a strong first-order PT at  tree-level entails an additional scaling relation between the effective couplings in the dimensionally reduced theory,
\begin{equation}
    m_{3}^2 T \sim \lambda_{3}T^2 \sim c_{\phi^6} T^3,
\end{equation}
with $m_3^2>0, \lambda_3<0, c_{\phi^6}>0$,
and the Higgs-field nucleation occurs at the soft scale $m_3^2\sim g^2 T^2$~\cite{Camargo-Molina:2024sde}.
This gives rise to the following scaling relation between the SM and BSM couplings:
\begin{equation}
\label{eq:couplings-scaling}
    \tilde{g}^2 \sim g \frac{\Lambda}{T}.
\end{equation}
Given this, we can establish the scalings of the effective couplings in \cref{eq:lagrangian3d};
for instance, the octic Higgs effective coupling parametrically is $c_{\phi^8} \sim g^3/\Lambda$~\footnote{The $c_{\phi^8}$ term could also be generated by $\bar{c}_{\phi^6}^2$, which parametrically is $\sim g^4/T$.}.

This establishes that the dimension-four operators in \cref{eq:lagrangian3d} feature an additional numerical suppression beyond the $1/T$ scaling from EFT counting. It is therefore consistent to restrict calculations to a single insertion of operators with mass dimension four.

\section{Counterterms and beta functions computation}
\label{sec:computation}
The main goal of this work is to renormalize the dimensionally reduced theory of \cref{eq:lagrangian3d} at the leading, two-loop order. 

We work in dimensional regularization (dimReg) with space-time dimension $d=3-2\epsilon$. The following observations simplify this task. (1) There are no logarithmic divergences at one loop, and therefore neither subdivergences nor double poles in two-loop diagrams. (2) Therefore, integrals can be IR regulated with a simple mass term,
even in the presence of gauge fields. (3) In 3D,
the only divergent integral is
\begin{equation}
\label{eq:UVintegral}
    \begin{split}
        &\int \frac{d^3 q}{(2\pi)^3}  \frac{d^3r}{(2\pi)^3} \frac{1}{(q^2+m_1^2) (r^2+m_2^2) ((q+r)^2+m_3^2)} \\&\stackrel{\text{dimReg}}{=} \frac{1}{64\pi^2\epsilon}+\frac{1}{16\pi^2}\left(\log\frac{\mu_{3D}}{m_1+m_2+m_3}+\frac{1}{2}\right) +\mathcal{O}(\epsilon) \,,
    \end{split}
\end{equation}
with $\mu_{3D}$ being the renormalization scale. The general form of this integral in $d$ dimensions can be found in Ref.~\cite{Davydychev:1992mt}. (4) Contrary to the 4D case, the number of relevant redundant operators is
significantly reduced (see \cref{eq:lagrangian3d}):
all others that exist in 4D, e.g., $D^2\phi^\dagger D^2\phi$ and $D_r W_{rs}^I (\phi^\dagger\overleftrightarrow{D_s}\sigma^I\phi)$, are of higher energy dimension in 3D. (5) Also, in contrast to the 4D case, gauge and quartic couplings are dimensionful. This, together with
the topological restriction $E=-2+\sum_i (i-2) n_i$---where $n_i$ stands for the valence of a given vertex $i$ and $E$ is the number of external legs---significantly constrains the relevant diagrams contributing to the renormalization.

\begin{figure*}[t]
    \includegraphics[width=2\columnwidth]{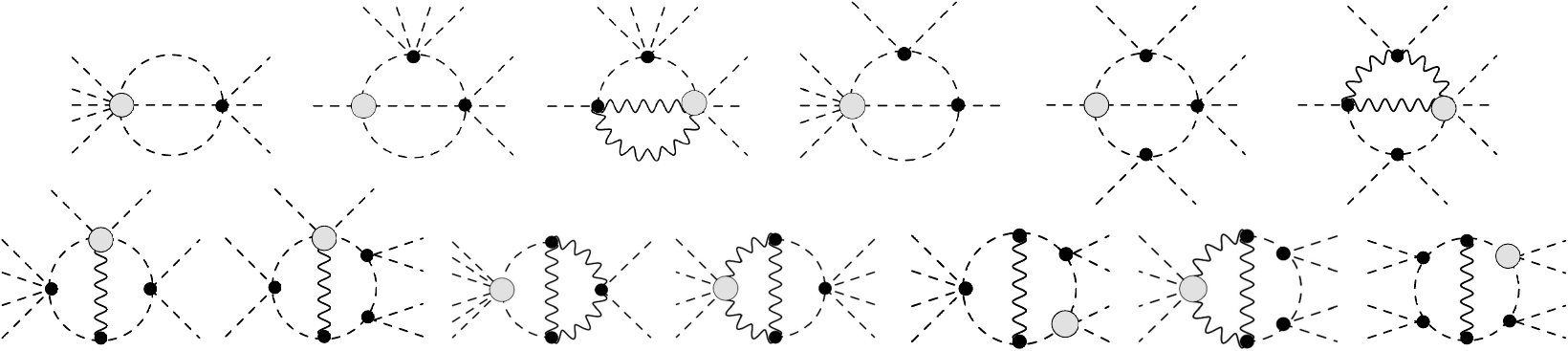}
    \caption{\it Example diagrams for the renormalization of $c_{\phi^8}$. Gray blobs represent effective operators of the second line of \cref{eq:lagrangian3d}, while black dots stand for renormalizable gauge, quartic and sextic scalar couplings.}
    \label{fig:diagrams}
\end{figure*}

 The last point is particularly useful to renormalize $c_{\phi^8}$ and $c_{\phi^6 W_0^2}$, since the number of two-loop diagrams with eight external legs is prohibitively large. For instance, 
 ignoring $W_0$ interactions, the topological constraint restricts the possible combinations of couplings entering the RG equations of $c_{\phi^8}$
 (with exactly one insertion of $c_{\phi^4D^2}, c_{\phi^8}, c_{\phi^2 W^2}$) to: $c_{\phi^6} c_{\phi_8}$, $c_{\phi^4 D^2} c_{\phi^6}^2$, $c_{\phi^4 D^2} c_{\phi^6} g_3^4$, $c_{\phi^8}\lambda_3^2$, $c_{\phi^4 D^2} \lambda_3^2 c_{\phi^6}$, $c_{\phi^4 D^2} g_3^6\lambda_3$, $c_{\phi^4 D^2} c_{\phi^6} g_3^2 \lambda_3$, $c_{\phi^4 D^2}g_3^2 \lambda_3^3$, $c_{\phi^8}g_3^4$, $c_{\phi^4 D^2} g_3^4 c_{\phi^6}$, $c_{\phi^4 D^2} c_{\phi^6} g_3^2 \lambda_3$, $c_{\phi^4 D^2}g_3^4\lambda_3^2$, $c_{\phi^4 D^2} g_3^2 \lambda_3^3$, $c_{\phi^2 W^2}c_{\phi^6}\lambda_3 g_3^2$, $c_{\phi^2 W^2}\lambda_3^3 g_3^2$, $c_{\phi^2 W^2}\lambda_3^2 g_3^4$, $c_{\phi^2 W^2}\lambda_3 g_3^6$, $c_{\phi^2 W^2}\lambda_3 g_3^6$ and $c_{\phi^2 W^2}g_3^8$;
representative diagrams are shown in ~\cref{fig:diagrams}. However, only the first two combinations result in the correct energy dimension, hence contributing to the $c_{\phi_8}$ running~\footnote{This argument can be further refined in fewer words. The coefficient $c_{\phi^8}$ can be only renormalized by loops involving effective operators, so either including $c_{\phi^8}$ itself, $c_{\phi^4 D^2}$ or $c_{\phi^2 W^2}$. The last one comes necessarily with powers of $g_3$, as it includes a gauge boson in the loop, which increases the total energy dimension. We therefore end up with only $c_{\phi^8} c_{\phi^6}$ and $c_{\phi^4D^2} c_{\phi^6}^2$.}.

Taking into account these considerations, we proceed by generating Feynman rules with \texttt{FeynRules}~\cite{Alloul:2013bka} and performing a suitable rotation to Euclidean space.
Subsequently, we evaluate two-loop $n$-point off-shell functions employing the background-field method~\cite{Abbott:1981ke}, using \texttt{FeynArts}~\cite{Hahn:2000kx} and \texttt{FeynCalc}~\cite{Shtabovenko:2023idz}, and expand the relevant two-loop integrals in the UV region, yielding the integrals of the form of \cref{eq:UVintegral}. 
This calculation results in the following two-loop counterterms in Feynman gauge in the \msbar-subtraction scheme, 

\begin{widetext}
\begin{equation}
\label{eq:counterterms}
    \begin{split}
        I^{-1}_{2} \delta Z_\phi  &= 
        2 c_{\phi^2 W^2} g_3^2 + c_{\phi^4D^2}^{(1)}\left(5  g_3^2- 8\lambda_3\right)- c_{\phi^4D^2}^{(2)} \left(\frac{3}{2}  g_3^2 - 2 \lambda_3\right) -2 c^{(1)}_{\phi^2 W_0^2 D^2}\lambda_{\phi^2W_0^2} + 4 c^{(2)}_{\phi^2 W_0^2 D^2} g_3^2\,\\
        I^{-1}_{2} \delta m_3^2 &= 
        -\frac{39}{16} g_3^4 - 9 g_3^2 \lambda_3 + 12 \lambda_3^2 - 12 g_3^2 \lambda_{\phi^2W_0^2} + 6 \lambda_{\phi^2W_0^2}^2\\
        &\quad - m_3^2\biggl[18 c_{\phi^2 W^2} g_3^2 + c_{\phi^4D^2}^{(1)} \left(9 g_3^2-24 \lambda_3  \right) - c_{\phi^4D^2}^{(2)} \left(\frac{3}{2}g_3^2  - 6 \lambda_3 \right) - 6 c^{(1)}_{\phi^2 W_0^2 D^2} \lambda_{\phi^2W_0^2}\biggr] \\
        &\quad+ m_D^2\biggl[-24 c_{\phi^2 W^2} g_3^2 + 12 c^{(1)}_{\phi^2 W_0^2 D^2} \left(g_3^2- \lambda_{\phi^2W_0^2}  \right) - 12 c^{(2)}_{\phi^2 W_0^2 D^2} g_3^2\biggr]
        \,, \\
        I^{-1}_{2} \delta \lambda_3 &= - c_{\phi^2 W^2}\left(\frac{39}{2} g_3^4 + 72 g_3^2 \lambda_3+ 48 g_3^2 \lambda_{\phi^2W_0^2} \right)  + c_{\phi^4D^2}^{(1)}\left(96 c_{\phi^6} m_3^2 +3 g_3^4 - 42 g_3^2 \lambda_3 +240 \lambda_3^2 + 24 \lambda_{\phi^2W_0^2}^2\right) \\
        &\quad - c_{\phi^4D^2}^{(2)}\left(24 c_{\phi^6} m_3^2 + \frac{13}{16} g_3^4 -6 g_3^2 \lambda_3 + 56 \lambda_3^2 + 6\lambda_{\phi^2W_0^2}^2 \right) 
        + c_{\phi^2 W_0^2 D^2}^{(2)} \left(\frac{3}{2} g_3^4 - 24 g_3^2 \lambda_{\phi^2 W_0^2}\right) \\
        &\quad + 4 c_{\phi^2 W_0^2 D^2}^{(1)} \left(c_{\phi^4 W_0^2}^{(1)} (3 m_3^2 - 6 m_D^2)+ c_{\phi^4 W_0^2}^{(2)} (m_3^2 - 2 m_D^2)+ \lambda_{\phi^2 W_0^2} (6 g_3^2 + 9 \lambda_3 - 12 \lambda_{\phi^2 W_0^2})\right) \\ 
        & \quad -   c_{\phi^6} \left(18 g_3^2 - 96  \lambda_3\right) +12 c_{\phi^4 W_0^2}^{(1)} (- g_3^2 + 2 \lambda_{\phi^2 W_0^2}) + 4 c_{\phi^4 W_0^2}^{(2)} (- g_3^2 + 2 \lambda_{\phi^2 W_0^2})
        \,,\\
        I^{-1}_{2} \delta c_{\phi^6} &=  204 c_{\phi^6}^2 + 24 (c_{\phi^4 W_0^2}^{(1)})^2+ 16 (c_{\phi^4 W_0^2}^{(2)})^2 +16 c_{\phi^4 W_0^2}^{(1)} c_{\phi^4 W_0^2}^{(2)} 
        \\
        &\quad + c_{\phi^6} \bigg[-162 c_{\phi^2 W^2} g_3^2 -c_{\phi^4D^2}^{(1)} \left(99  g_3^2 - 1800 \lambda_3 \right) +c_{\phi^4D^2}^{(2)} \left(\frac{27}{2}  g_3^2 - 426  \lambda_3 \right) \bigg]  \\
        &\quad 
        +8c_{\phi^4 W_0^2}^{(1)} \bigg[3 c_{\phi^2 W_0^2 D^2}^{(1)} (g_3^2 + 3 \lambda_3 - 7 \lambda_{\phi^2 W_0^2}) - 3 c_{\phi^2 W_0^2 D^2}^{(2)} g_3^2 - 6 c_{\phi^2 W^2} g_3^2 + 3 \lambda_{\phi^2 W_0^2} (4 c_{\phi^4 D^2}^{(1)} - c_{\phi^4 D^2}^{(2)}) \bigg]  \\
        &\quad 
        +8 c_{\phi^4 W_0^2}^{(2)} \bigg[c_{\phi^2 W_0^2 D^2}^{(1)} (g_2^2 + 3 \lambda_3 - 7 \lambda_{\phi^2 W_0^2}) - c_{\phi^2 W_0^2 D^2}^{(2)} g_2^2 - 2 c_{\phi^2 W^2} g_2^2+  \lambda_{\phi^2 W_0^2} (4 c_{\phi^4 D^2}^{(1)} - c_{\phi^4 D^2}^{(2)})\bigg]  \\
        &\quad 
        - 30 c_{\phi^8} \left(g_3^2 - 8 \lambda_3\right) - 12 c_{\phi^6 W_0^2}^{(1)} (g_2^2 - 3 \lambda_{\phi^2 W_0^2}) + 4 c_{\phi^6 W_0^2}^{(2)} (g_2^2 - 3\lambda_{\phi^2 W_0^2}) 
        \, \\
        I^{-1}_{2} \delta c_{\phi^8} &=  12 c_{\phi^6} \bigg[288 c_{\phi^4D^2}^{(1)} c_{\phi^6} - 69 c_{\phi^4D^2}^{(2)} c_{\phi^6}+ 5 c_{\phi^2 W_0^2 D^2}^{(1)} ( 3 c_{\phi^4 W_0^2}^{(1)} + c_{\phi^4 W_0^2}^{(2)}) +88  c_{\phi^8} \bigg] 
        \\
        &\quad -8 c_{\phi^4 W_0^2}^{(1)} \bigg[ 6 c_{\phi^2 W_0^2 D^2}^{(1)} (3 c_{\phi^4 W_0^2}^{(1)} + 2 c_{\phi^4 W_0^2}^{(2)}) - 4 c_{\phi^4 D^2}^{(1)} (3 c_{\phi^4 W_0^2}^{(1)} + 2 c_{\phi^4 W_0^2}^{(2)})  \\
        &\quad + c_{\phi^4 D^2}^{(2)} (3 c_{\phi^4 W_0^2}^{(1)} + 2 c_{\phi^4 W_0^2}^{(2)}) - 9 c_{\phi^6 W_0^2}^{(1)} - 3 c_{\phi^6 W_0^2}^{(2)}\bigg]
        \\
        &\quad
        -8 c_{\phi^4 W_0^2}^{(2)} \bigg[ 2 c_{\phi^4 W_0^2}^{(2)} (5 c_{\phi^2 W_0^2 D^2}^{(1)}- c_{\phi^2 W_0^2 D^2}^{(2)}) - c_{\phi^4 W_0^2}^{(2)} (4 c_{\phi^4 D^2}^{(1)} - c_{\phi^4 D^2}^{(2)}) - 3 c_{\phi^6 W_0^2}^{(1)} - 5 c_{\phi^6 W_0^2}^{(2)} \bigg] \,, \\
        I^{-1}_{2} \delta c_{\phi^4D^2}^{(1)} &= 
        4 c_{\phi^6}\left(4c_{\phi^4D^2}^{(1)} - c_{\phi^4D^2}^{(2)}\right) + 2 c_{\phi^2 W_0^2 D^2}^{(1)} \left(c_{\phi^4 W_0^2}^{(1)} + \frac{1}{3}c_{\phi^4 W_0^2}^{(2)}\right)
        \,, \\
        I^{-1}_{2} \delta c_{\phi^4D^2}^{(2)} &= 0\,,\\
        I^{-1}_{2} \delta r_{\phi^4D^2}^{(3)} &= -8 c_{\phi^6}\left(4c_{\phi^4D^2}^{(1)} - c_{\phi^4D^2}^{(2)}\right) - 4 c_{\phi^2 W_0^2 D^2}^{(1)} \left(c_{\phi^4 W_0^2}^{(1)} + \frac{1}{3}c_{\phi^4 W_0^2}^{(2)}\right) \,,\\
    \end{split}
\end{equation}
\end{widetext}
where we have factored out $I_2^{-1} = 64\pi^2\epsilon$ ---the two-loop \msbar\, $\epsilon$ pole of \cref{eq:UVintegral}. We only explicitly show counterterms for Higgs-only operators, 
whereas the rest of the counterterms can
be found in the Supplemental Material attached.

The super-renormalizable part of the $\delta m_3^2$ and $\delta m_D^2$ counterterms agrees with the results in Ref.~\cite{Laine:1995np}, which differ by a factor of $2$ in $\delta m_D^2$ with respect to Ref.~\cite{Farakos:1994kx}.

We proceed by canonically normalizing the kinetic terms of the scalar and gauge fields and removing the redundant operators
$\mathcal{O}_{\phi^4 D^2}^{(3)}$, $\mathcal{O}_{\phi^2 W_0^2 D^2}^{(4)}$ and $\mathcal{O}_{\phi^2 W_0^2 D^2}^{(5)}$, by redefining
$\phi\to \phi -\frac{1}{2}r_{\phi^4 D^2}^{(3)}\phi (\phi^\dagger\phi) + r_{\phi^2 W_0^2 D^2}^{(4)}\phi W_0^I W_0^I $, $W_0^I \to W_0^I + r_{\phi^2 W_0^2 D^2}^{(5)}|\phi|^2W_0^I$.
This results in the following shifts for the counterterms:
\begin{widetext}
\begin{equation}
    \begin{split}
        \delta m_3^{\prime 2}&=\delta m_3^2 - \delta Z_\phi m_3^2\,, \\
        \delta m_D^{\prime 2}&=\delta m_D^2 - \delta Z_{W_0} m_D^2\,, \\
        \delta\lambda_3'&=\delta\lambda_3 - 2 \delta Z_\phi\lambda_3- m_3^2 \delta r_{\phi^4 D^2}^{(3)}\,,\\
        \delta\lambda_{\phi^2 W_0^2}'&=\delta\lambda_{\phi^2 W_0^2} - (\delta Z_\phi+\delta Z_{W_0}) \lambda_{\phi^2 W_0^2}+ m_3^2 \delta r_{\phi^2 W_0^2 D^2}^{(4)}+ m_D^2 \delta r_{\phi^2 W_0^2 D^2}^{(5)}\,,\\
        \delta\lambda_{W_0^4}'&=\delta\lambda_{W_0^4} - 2 \delta Z_{W_0} \lambda_{W_0^4},\\
        \delta c_{\phi^6}'&=\delta c_{\phi^6} - 3 \delta Z_\phi c_{\phi^6} - 2\lambda_3 \delta r_{\phi^4 D^2}^{(3)}\,, \\
        \delta c_{\phi^2 W_0^4}'&=\delta c_{\phi^2 W_0^4} - (\delta Z_\phi+2\delta Z_{W_0}) c_{\phi^2 W_0^4}+2\lambda_{\phi^2 W_0^2} \delta r_{\phi^2 W_0^2 D^2}^{(4)}+4\lambda_{W_0^4} \delta r_{\phi^2 W_0^2 D^2}^{(5)}\,, \\
        \delta c_{\phi^4 W_0^2}^{(1)\prime}&=\delta c_{\phi^4 W_0^2}^{(1)} - (2\delta Z_\phi+\delta Z_{W_0})  c_{\phi^4 W_0^2}^{(1)}+4\lambda_{3} \delta r_{\phi^2 W_0^2 D^2}^{(4)}+2\lambda_{\phi^2 W_0^2} \delta r_{\phi^2 W_0^2 D^2}^{(5)}\,, \\
        \delta c_{\phi^4 W_0^2}^{(2)\prime}&=\delta c_{\phi^4 W_0^2}^{(2)} - (2\delta Z_\phi+\delta Z_{W_0}) c_{\phi^4 W_0^2}^{(2)}\,, \\
        \delta c_{W_0^6}'&=\delta c_{W_0^6} - 3 \delta Z_{W_0} c_{W_0^6}\,, \\
        \delta c_{\phi^4D^2}^{(1)\prime}&=\delta c_{\phi^4D^2}^{(1)}   + \frac{1}{2}\delta r_{\phi^4 D^2}^{(3)}\,,\\
        \delta c_{\phi^4D^2}^{(2)\prime}&= \delta c_{\phi^4D^2}^{(2)}\,,\\
        \delta c_{\phi^8}' &= \delta c_{\phi^8} - 3c_{\phi^6}\delta r_{\phi^4 D^2}^{(3)}\,,\\
        \delta c_{\phi^2W^2}'&= \delta c_{\phi^2W^2}\,\\
        \delta c_{\phi^6 W_0^2}^{(1) \prime}&=\delta c_{\phi^6 W_0^2}^{(1)} + 2 c_{\phi^4 W_0^2}^{(1)}\delta r_{\phi^2 W_0^2 D^2}^{(5)} + 6 c_{\phi^6}\delta r_{\phi^2 W_0^2 D^2}^{(4)}  \,, \\
        \delta c_{\phi^6 W_0^2}^{(2)\prime}&=\delta c_{\phi^6 W_0^2}^{(2)} +2  c_{\phi^4 W_0^2}^{(2)}\delta r_{\phi^2 W_0^2 D^2}^{(5)} \,, \\
        \delta c_{\phi^2 W_0^2 D^2}^{(1),(2),\prime}&=\delta c_{\phi^2 W_0^2 D^2}^{(1),(2)}\,,
    \end{split}
\end{equation}
\end{widetext}
in which we only kept counterterms with up to one insertion of dimension-four operators.
Following these shifts, the running of the \msbar parameters is determined simply by~\cite{Fonseca:2025zjb},
\begin{equation}
    \beta_{c} \equiv 16\pi^2\mu \frac{dc}{d\mu} = 64 \pi^2 \epsilon \delta c\,,
\end{equation}
and for the effective couplings under scope, reads,
\begin{widetext}
\begin{equation}
    \label{eq:betafunctions}
        \begin{split}
        \beta_{m_3^2} &=   \overbrace{
        -\frac{39}{16} g_3^4 - 9 g_3^2 \lambda_3 + 12 \lambda_3^2 - 12 g_3^2 \lambda_{\phi^2W_0^2} + 6 \lambda_{\phi^2W_0^2}^2}^{\beta_{m_3^2}^{\text{LO}}}\\
        &\quad - m_3^2\biggl[20 c_{\phi^2 W^2} g_3^2 + c_{\phi^4D^2}^{(1)} \left(14 g_3^2-32 \lambda_3  \right) - c_{\phi^4D^2}^{(2)} \left(3 g_3^2  - 8 \lambda_3 \right) - 8 c^{(1)}_{\phi^2 W_0^2 D^2} \lambda_{\phi^2W_0^2} + 4 c^{(2)}_{\phi^2 W_0^2 D^2} g_3^2\biggr] \\
        &\quad+ m_D^2\biggl[-24 c_{\phi^2 W^2} g_3^2 + 12 c^{(1)}_{\phi^2 W_0^2 D^2} \left(g_3^2- \lambda_{\phi^2W_0^2}  \right) - 12 c^{(2)}_{\phi^2 W_0^2 D^2} g_3^2\biggr]
        \,, \\
        \beta_{\lambda_3} &= - c_{\phi^2 W^2}\left(\frac{39}{2} g_3^4 + 76 g_3^2 \lambda_3+ 48 g_3^2 \lambda_{\phi^2W_0^2} \right)  + c_{\phi^4D^2}^{(1)}\left(128 c_{\phi^6} m_3^2 +3 g_3^4 - 52 g_3^2 \lambda_3 +256 \lambda_3^2 + 24 \lambda_{\phi^2W_0^2}^2\right) \\
        &\quad - c_{\phi^4D^2}^{(2)}\left(32 c_{\phi^6} m_3^2 + \frac{13}{16} g_3^4 -9 g_3^2 \lambda_3 + 60 \lambda_3^2 + 6\lambda_{\phi^2W_0^2}^2 \right) 
        + c_{\phi^2 W_0^2 D^2}^{(2)} \left(\frac{3}{2} g_3^4 - 24 g_3^2 \lambda_{\phi^2 W_0^2} -8 g_3^2 \lambda_3 \right) \\
        &\quad + 4 c_{\phi^2 W_0^2 D^2}^{(1)} \left(c_{\phi^4 W_0^2}^{(1)} (4 m_3^2 - 6 m_D^2)+ c_{\phi^4 W_0^2}^{(2)} (\frac{4}{3} m_3^2 - 2 m_D^2)+ \lambda_{\phi^2 W_0^2} (6 g_3^2 + 10 \lambda_3 - 12 \lambda_{\phi^2 W_0^2})\right) \\ 
        & \quad -   c_{\phi^6} \left(18 g_3^2 - 96  \lambda_3\right) +12 c_{\phi^4 W_0^2}^{(1)} (- g_3^2 + 2 \lambda_{\phi^2 W_0^2}) + 4 c_{\phi^4 W_0^2}^{(2)} (- g_3^2 + 2 \lambda_{\phi^2 W_0^2})
        \,,\\
        \beta_{c_{\phi^6}} &=  204 c_{\phi^6}^2 + 24 (c_{\phi^4 W_0^2}^{(1)})^2+ 16 (c_{\phi^4 W_0^2}^{(2)})^2 +16 c_{\phi^4 W_0^2}^{(1)} c_{\phi^4 W_0^2}^{(2)} 
        \\
        &\quad + 2 c_{\phi^6} \bigg[-84 c_{\phi^2 W^2} g_3^2 -c_{\phi^4D^2}^{(1)} \left(57  g_3^2 - 944 \lambda_3 \right) +c_{\phi^4D^2}^{(2)} \left(9  g_3^2 - 224  \lambda_3 \right) + 3 c_{\phi^2 W_0^2 D^2}^{(1)} \lambda_{\phi^2 W_0^2} - 6 c_{\phi^2 W_0^2 D^2}^{(2)} g_3^2 \bigg]  \\
        &\quad 
        +8c_{\phi^4 W_0^2}^{(1)} \bigg[3 c_{\phi^2 W_0^2 D^2}^{(1)} (g_3^2 + 4 \lambda_3 - 7 \lambda_{\phi^2 W_0^2}) - 3 c_{\phi^2 W_0^2 D^2}^{(2)} g_3^2 - 6 c_{\phi^2 W^2} g_3^2 + 3 \lambda_{\phi^2 W_0^2} (4 c_{\phi^4 D^2}^{(1)} - c_{\phi^4 D^2}^{(2)}) \bigg]  \\
        &\quad 
        +8 c_{\phi^4 W_0^2}^{(2)} \bigg[c_{\phi^2 W_0^2 D^2}^{(1)} (g_2^2 + \frac{4}{3} \lambda_3 - 7 \lambda_{\phi^2 W_0^2}) - c_{\phi^2 W_0^2 D^2}^{(2)} g_2^2 - 2 c_{\phi^2 W^2} g_2^2+  \lambda_{\phi^2 W_0^2} (4 c_{\phi^4 D^2}^{(1)} - c_{\phi^4 D^2}^{(2)})\bigg]  \\
        &\quad 
        - 30 c_{\phi^8} \left(g_3^2 - 8 \lambda_3\right) - 12 c_{\phi^6 W_0^2}^{(1)} (g_2^2 - 3 \lambda_{\phi^2 W_0^2}) + 4 c_{\phi^6 W_0^2}^{(2)} (g_2^2 - 3\lambda_{\phi^2 W_0^2}) 
        \, \\
        \beta_{c_{\phi^8}} &=  12 c_{\phi^6} \bigg[296 c_{\phi^4D^2}^{(1)} c_{\phi^6} - 71 c_{\phi^4D^2}^{(2)} c_{\phi^6}+  16c_{\phi^2 W_0^2 D^2}^{(1)} ( c_{\phi^4 W_0^2}^{(1)} + \frac{1}{3} c_{\phi^4 W_0^2}^{(2)}) +88  c_{\phi^8} \bigg] 
        \\
        &\quad -8 c_{\phi^4 W_0^2}^{(1)} \bigg[ 6 c_{\phi^2 W_0^2 D^2}^{(1)} (3 c_{\phi^4 W_0^2}^{(1)} + 2 c_{\phi^4 W_0^2}^{(2)}) - 4 c_{\phi^4 D^2}^{(1)} (3 c_{\phi^4 W_0^2}^{(1)} + 2 c_{\phi^4 W_0^2}^{(2)})  \\
        &\quad + c_{\phi^4 D^2}^{(2)} (3 c_{\phi^4 W_0^2}^{(1)} + 2 c_{\phi^4 W_0^2}^{(2)}) - 9 c_{\phi^6 W_0^2}^{(1)} - 3 c_{\phi^6 W_0^2}^{(2)}\bigg]
        \\
        &\quad
        -8 c_{\phi^4 W_0^2}^{(2)} \bigg[ 2 c_{\phi^4 W_0^2}^{(2)} (5 c_{\phi^2 W_0^2 D^2}^{(1)}- c_{\phi^2 W_0^2 D^2}^{(2)}) - c_{\phi^4 W_0^2}^{(2)} (4 c_{\phi^4 D^2}^{(1)} - c_{\phi^4 D^2}^{(2)}) - 3 c_{\phi^6 W_0^2}^{(1)} - 5 c_{\phi^6 W_0^2}^{(2)} \bigg]
        \,,
        \end{split}
\end{equation}
\end{widetext}

The first line of $\beta_{m_3^2}$ in \cref{eq:betafunctions} is the exact running of the mass parameter, $m_3^2$, if we were to truncate the 3D Lagrangian in \cref{eq:lagrangian3d} to only include super-renormalizable operators~\cite{Farakos:1994kx}, while the rest of the running is triggered by the remaining effective couplings. 
The quartic, $\lambda_3$, and the sextic, $c_{\phi^6}$, scalar operators get renormalized with the inclusion of the dimension-three operators.

The remaining beta functions for the effective couplings are provided in the Supplemental Material attached.
We remark that, in the physical basis, none of the effective couplings in
the class $c_{\phi^4D^2}$, nor $c_{\phi^2 W_0^2 D^2}$ run at the two-loop level with up to one insertion of the dimension-four operators. 

As a cross-check, we have verified that the running of $m_3^2$, $\lambda_3$, $c_{\phi_6}$, and $c_{\phi_8}$ in \cref{eq:betafunctions}, neglecting operators with derivatives, precisely reproduces the logarithmic dependence, $\log(\mu_{3D})$, of the two-loop effective potential, explicitly showing the better control of the renormalization-scale dependence allowed by the derived RGEs.

\begin{figure}[b]
    \centering
    \includegraphics[width=\linewidth]{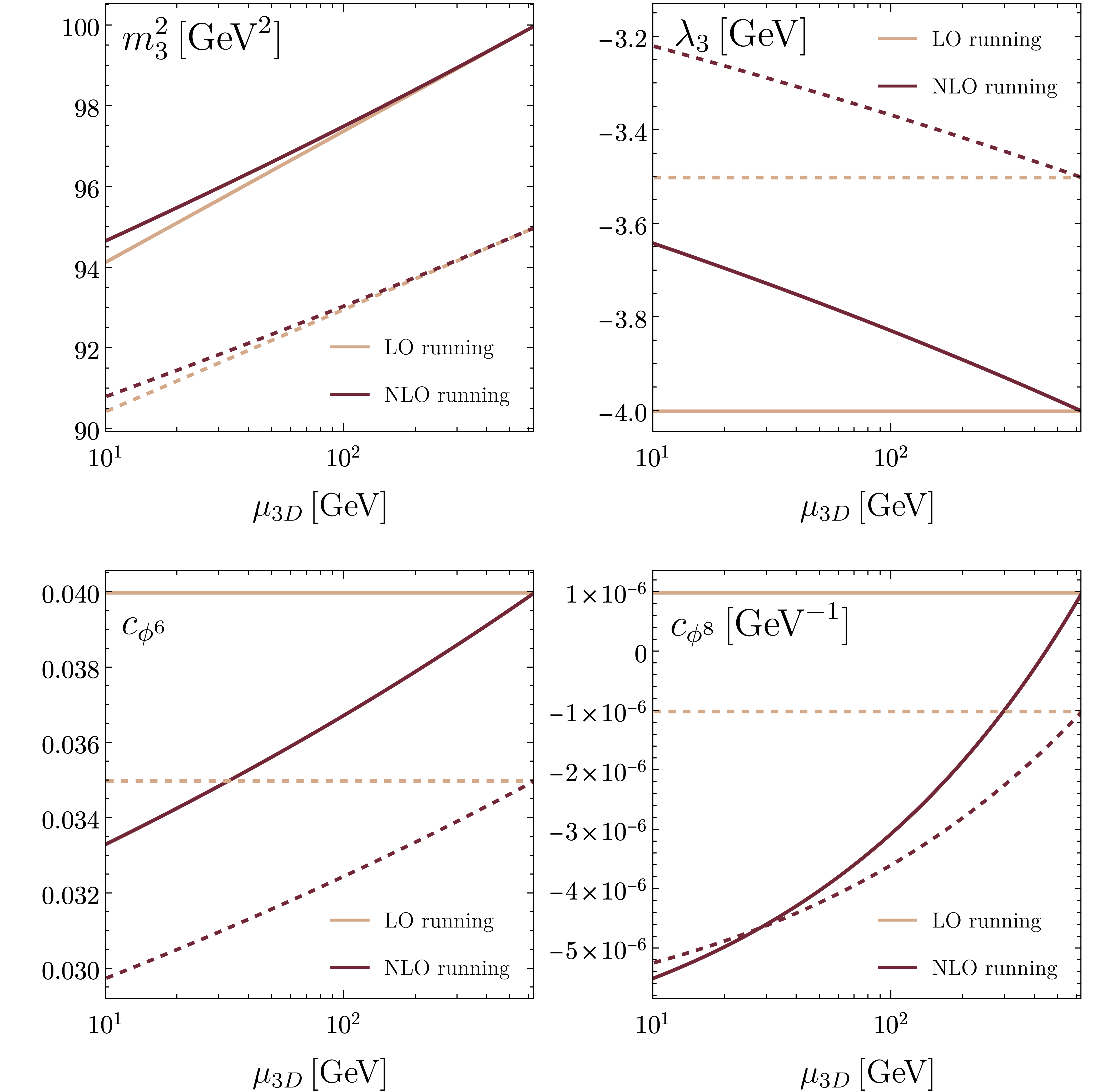}
    \caption{\it Running of the effective couplings $m_3^2$, $\lambda_3$, $c_{\phi^6}$ and $c_{\phi^8}$ from the hard scale $2\pi T$ down to $\mu_{3D}$.
    We assume
    $T = 100\,\GeV$, and take the effective couplings $\{m^2_{3},m_D^2,\lambda_{3}, \lambda_{W_0^4}, c_{\phi^6}, c_{\phi^4D^2}^{(1)}, c_{\phi^4D^2}^{(2)}, c_{\phi^8}, g_3^2 \}$ to be $\{100,100,-4, 0.01, 0.04, 10^{-4}, 10^{-4}, 10^{-6}, 2 \}$ (solid) and $\{95,100,-3.5, 0.01, 0.035, 10^{-4}, 10^{-4}, -10^{-6}, 2 \}$ (dashed) with all values given in $\GeV$ at the hard scale, with the rest of effective coupling taken to be zero.}
    \label{fig:runningplot}
\end{figure}

\section{Scalar potential modification}
\label{sec:scalarpotential}
The significance of the running of the effective parameters can be estimated based on the power counting introduced in \cref{sec:dimredtheory}, in case the tree-level barrier is present:
\begin{equation}
\label{eq:runningscaling}
    \begin{split}
        \frac{\partial \log m_3^2}{\partial \log \mu_{3D}} &\sim \overbrace{g^2}^{\text{LO}} + g^3 \frac{T}{\Lambda}\, , \\
        \frac{\partial \log \lambda_3}{\partial \log \mu_{3D}} &\sim \frac{\partial \log c_{\phi^6}}{\partial \log \mu_{3D}} \sim \frac{\partial \log c_{\phi^8}}{\partial \log \mu_{3D}} \sim g^2\,, \\
        \frac{\partial \log g_3^2}{\partial \log \mu_{3D}} &\sim g^5 \frac{T}{\Lambda}\,.
    \end{split} 
\end{equation}
It indicates that the LO $m_3^2$ running---triggered by the super-renormalizable operators---is parametrically of the same order as the quartic, sextic, and octic running. 

To illustrate how these couplings run, we estimate the values of the Wilson coefficients according to \cref{eq:all-scalings-soft} and assume $g = 0.1$, $\Lambda = 1\,\text{TeV}$ and $T = 100\,\GeV$.
Moreover, for simplicity, through this section, we assume that $\lambda_{\phi^2 W_0^2}=g^2_3/4$, which holds up to higher-order corrections~\cite{Farakos:1994kx}.
We numerically solve the coupled RG equations and present the benchmark running from the hard scale ($2\pi T$) down to $\mu_{3D}$ in \cref{fig:runningplot}.

As evident from \cref{fig:runningplot}, the effective operators trigger the running of $\lambda_3$, $c_{\phi^6}$, resulting in changes of their values at the level of ten percent, while the correction due to next-to-leading order (NLO) $m_3^2$ running is on a percent level. Notably, for the chosen benchmark points, $c_{\phi^8}$ changes significantly, flipping its sign as it evolves toward small scales; the running of $c_{\phi^8}$ is additionally enhanced by large numerical values in its beta function compared to the estimate in \cref{eq:runningscaling}. 
Although we present only two reference benchmark points, we find that this behavior is universal within the
parameter space corresponding to tree-level barrier nucleation, except for cases where tree-level–enhanced dimension-four operators are present.

The impact of the running on the potential is discussed further below.

\begin{figure}[b]
    \centering
    \includegraphics[width=\linewidth]{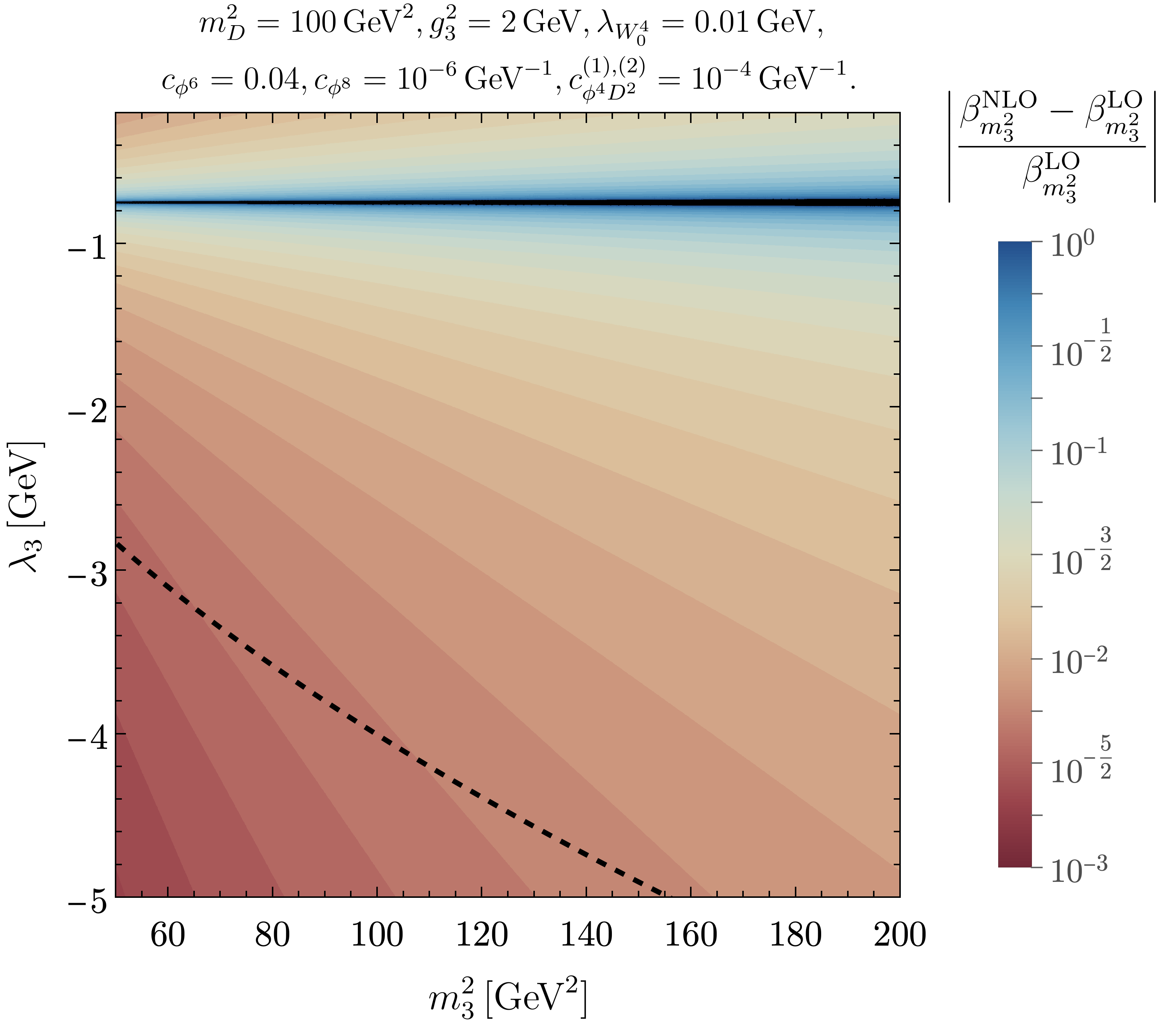}
    \caption{\it Relative correction to the $m_3^2$ beta function from the effective operators. The black dashed line denotes the parameter values for which two minima of the scalar potential are degenerate. 
    }
    \label{fig:m3betaratio}
\end{figure}

Next, we assess the numerical relevance of the NLO correction to the running of the mass parameter, $m_3^2$.
In \cref{fig:m3betaratio}, we plot the relative correction to the $m_3^2$ beta function from the effective couplings, by defining $\beta_{m_3^2}^{\text{LO}}$---the super-renormalizable part of $m_3^2$ running--- and $\beta_{m_3^2}^{\text{NLO}}$--- the full two-loop running of $m_3^2$---per \cref{eq:betafunctions}. The black dashed line represents the values of the parameters for which two local minima of the potential become degenerate in energy (this corresponds to $T$ being the critical temperature); therefore, below this line, the broken minimum is the global minimum of the scalar potential.
The relative correction to the $m_3^2$ running is small (typically on a percent level) for the wide range of parameter space; however, in the vicinity of the region given by $\lambda_3 = -3g_3^2/8$  (taking $\lambda_{\phi^2 W_0^2}=g^2_3/4$ as discussed above), where the LO running vanishes, i.e. where $\beta_{m_3^2}^{\text{LO}}=0$), corrections from the effective operators to the beta function become dominant. 

\begin{figure}[t]
    \centering
    \includegraphics[width=\linewidth]{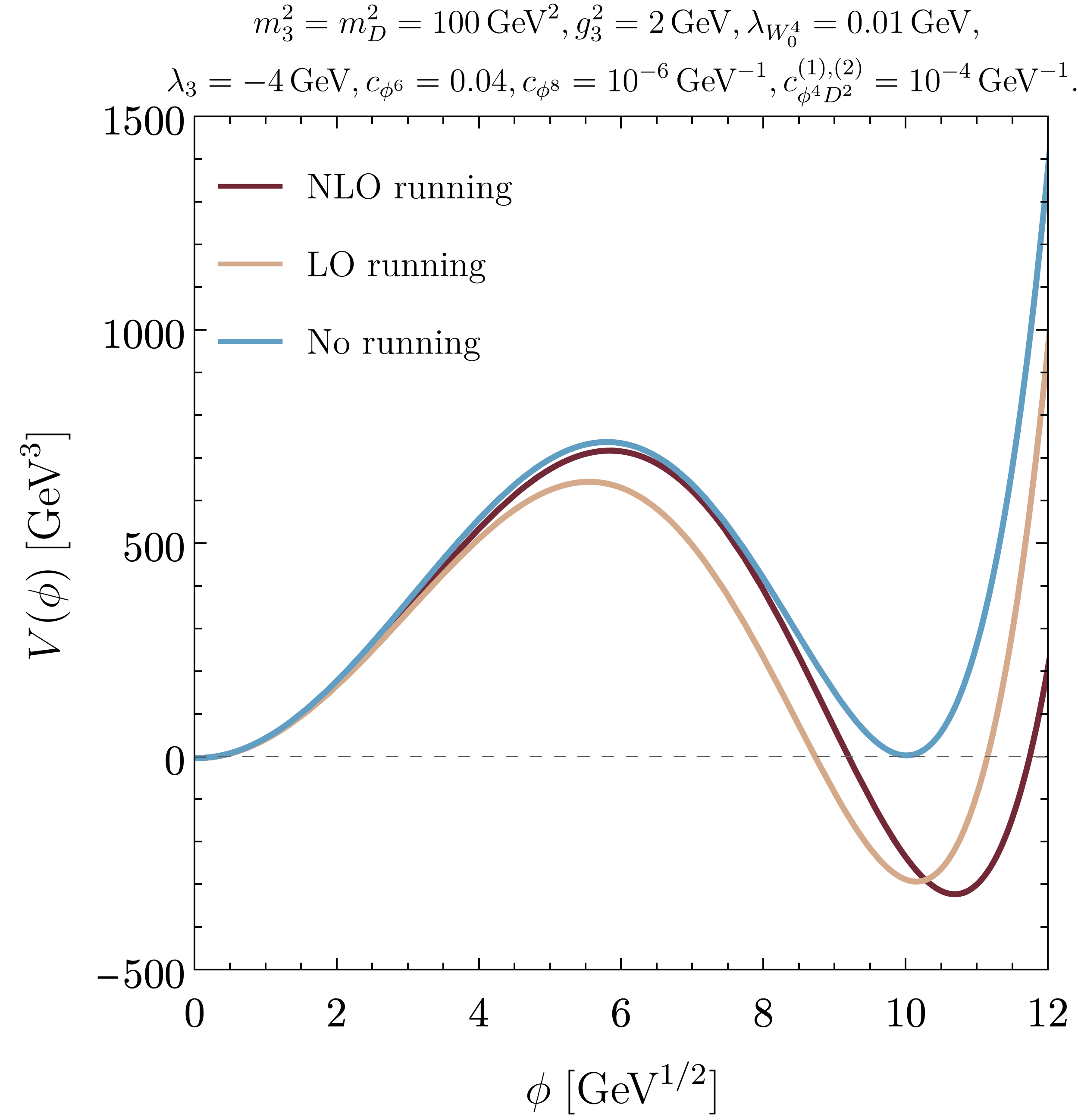}
    \caption{\it Scalar potential $V (\phi)$ without running, with LO running, and with NLO running of the effective couplings. Parameters run from $2\pi T$ down to $g T$, with $g=0.1$ and $T=100$ GeV.
    }
    \label{fig:potential}
\end{figure}

In \cref{fig:potential}, the shape of the scalar potential is shown with and without the inclusion of the running of the effective parameters. 
As evident from the figure, contributions from the effective couplings can substantially alter the shape of the scalar potential and shift the position of the broken minimum.
The main difference between including only the LO mass running and the NLO running with effective operators arises from the running of the quartic and sextic scalar couplings, $\lambda_3$ and $c_{\phi^6}$, respectively; additional terms in the mass parameter running have only a minor effect on the scalar potential. 
It is worth recalling that the inclusion of the running to the scalar potential can be seen as resumming higher effective potential loops.

\begin{figure*}[t]
    \centering
    \includegraphics[width=\linewidth]{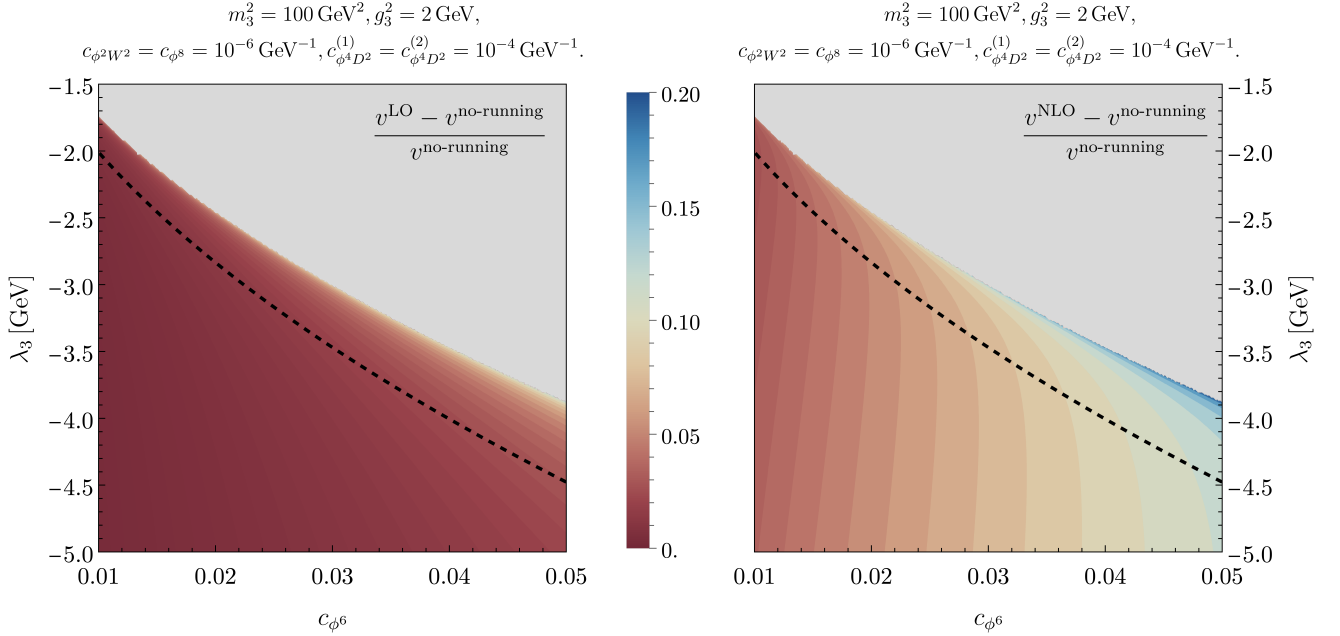}
    \caption{\it Relative change of the position of the broken minimum of the scalar potential ($v$) with the inclusion of the LO mass parameter running only \textit{(left)} and NLO running of the effective parameters \textit{(right)}. Parameters run from $2\pi T$ down to $g T$, with $g=0.1$ and $T=100$ GeV. Grey regions represent the parameter space where the broken minimum is absent in case no running is included. The black-dashed line denotes the parameter values for which two minima of the potential are degenerate in the absence of RG running. 
    }
    \label{fig:deltavev}
\end{figure*}

We further investigate the impact of the running by studying the shift in the broken minimum field value $v$. 
This is a key parameter, since 
the ratio $\xi = v/T$ 
characterizes the baryon washout during the baryogenesis at the first-order PT~\cite{Patel:2011th}. (Note, however, that beyond tree level, $\xi$ is gauge dependent~\cite{Patel:2011th}.)

In \cref{fig:deltavev}, the shift in this quantity caused by including the running of the effective parameters is illustrated. The \textit{left} (\textit{right}) panel shows the shift when only the LO running of the mass parameter (NLO running of all operators) is taken into account.
Including the NLO running of the effective parameters can alter the broken field value by tens of percent, while the LO running (where only $m_3^2$ runs) produces a much smaller shift, at the few-percent level. 
The difference between LO and NLO running becomes even more prominent with the increase of the effective couplings, e.g., $c_{\phi^4D^2}$.

As before, the black-dashed line in \cref{fig:deltavev} represents the parameter values for which the broken and unbroken minima are degenerate. 
Below this line, the broken minimum becomes the global minimum, and it is within this region that PT nucleation occurs; hence, it is of main interest to the PT phenomenology.
In this region, the LO running can only lead to a few percent shift to the broken minimum value, while the NLO running of the effective couplings could result in changes on the order of tens of percent, specifically for large values of the sextic coupling $c_{\phi^6}$.

\begin{figure}[t]
    \centering
    \includegraphics[width=\linewidth]{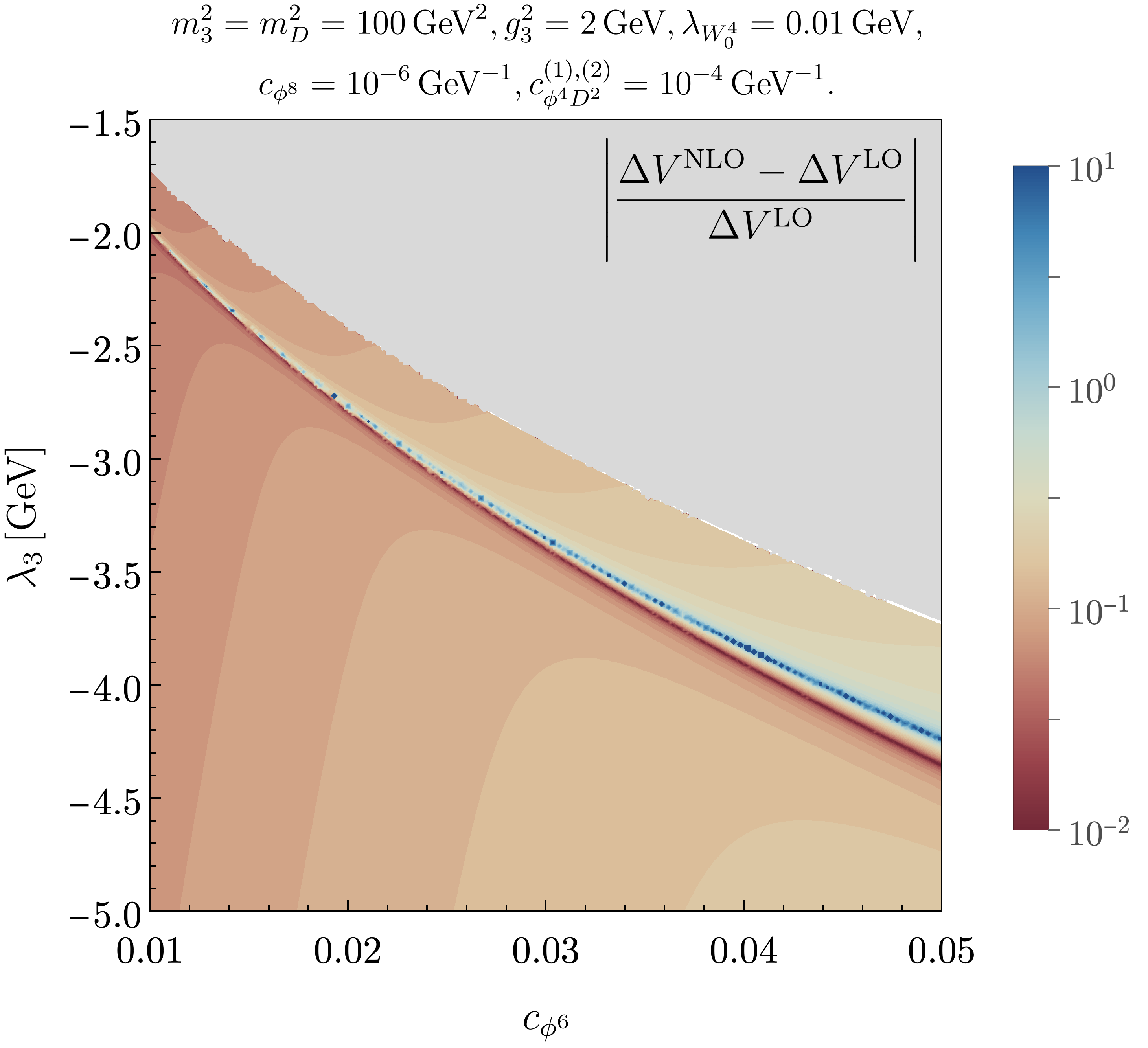}
    \caption{\it Relative shift of the potential difference $\Delta V$ between broken and unbroken minima between cases of inclusion of LO and NLO running of the effective parameters. Parameters run from $2\pi T$ down to $g T$, with $g=0.1$ and $T=100$ GeV. The grey region represents the parameter space where the broken minimum is absent in case the NLO running is included.
    }
    \label{fig:deltaV}
\end{figure}

Similarly, the NLO correction to the running of the effective parameters affects the difference in the potential values between two local minima, which in turn determines the strength of the PT. 
In \cref{fig:deltaV}, we present the relative change in the potential difference $\Delta V$ between the broken and unbroken minima when comparing the inclusion of LO and NLO running of the effective parameters.
As can be seen from the figure, the potential difference can change by up to tens of percent, depending on the specific parameter point, therefore modifying the strength of the PT.
The blue diagonal region simply points to the values of the couplings where $\Delta V^{\text{LO}}$ crossed zero value, while $\Delta V^{\text{LO}} =\Delta V^{\text{NLO}}$ holds in the dark red region.
We note that for points where two local minima are of similar depth, the NLO corrections lift the broken minima, while for points where the broken minimum is significantly deeper than the unbroken, NLO corrections deepen it, resulting in stronger PTs.

\begin{figure}[t]
    \centering
    \includegraphics[width=\linewidth]{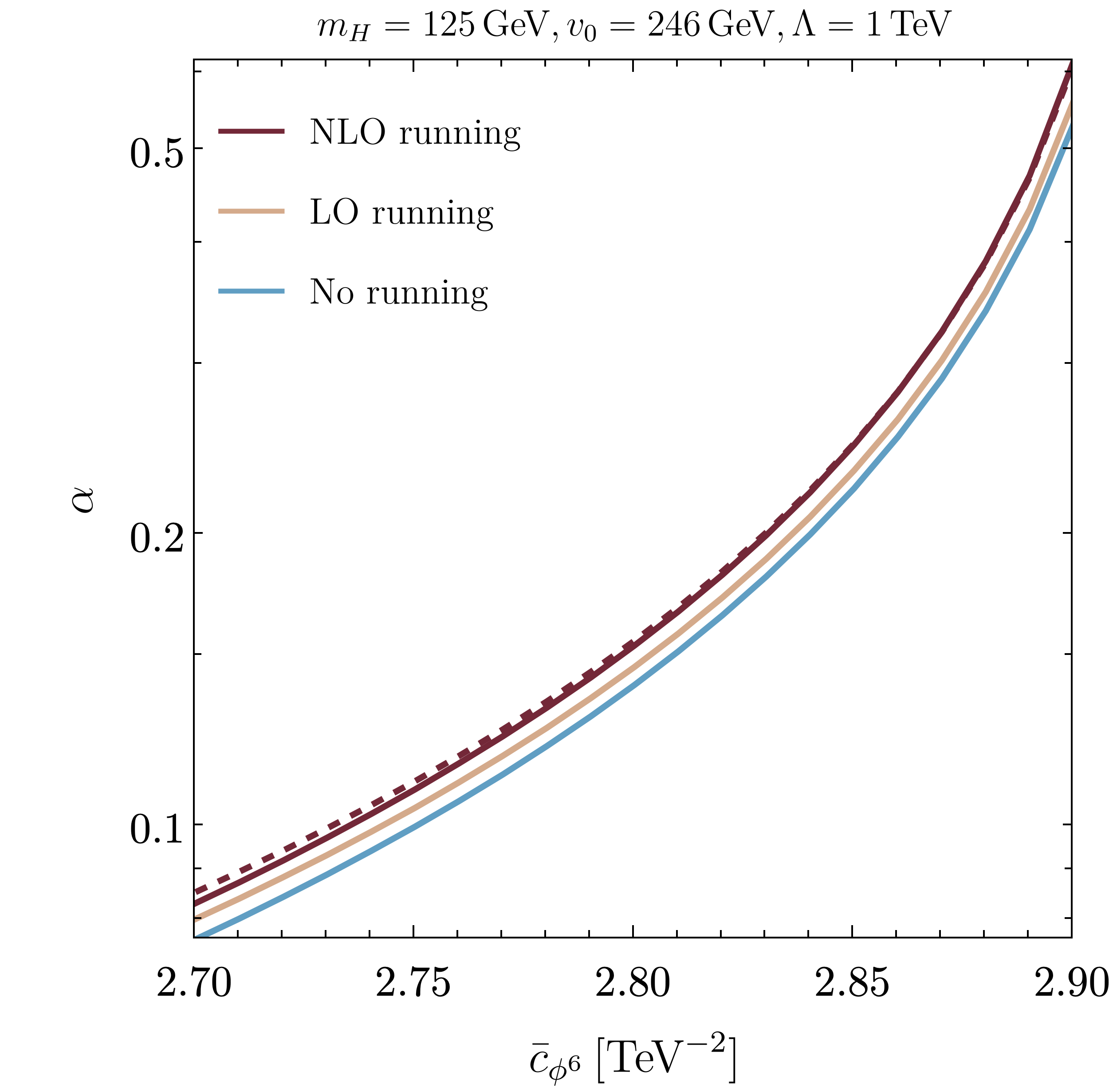}
    \caption{\it PT strength $\alpha$ determined without and with the running of the effective parameters from the hard thermal scale $2\pi T$ down to $m_3$ scale. The dashed line with NLO running corresponds to taking $\bar{c}^{(1)}_{\phi^4D^2}$ to be negative. 
    }
    \label{fig:alpha}
\end{figure}

For completeness, in \cref{fig:alpha}, we present the effect of the running of the effective parameters on the strength of the PT $\alpha$ given by~\cite{Caprini:2019egz,Caprini:2024hue}
\begin{equation}
    \alpha = \frac{1}
    {\rho_{\text{rad}}}\left(-\frac{3}{4}T \Delta V+\frac{1}{4}T^2\frac{d\Delta V}{dT}\right)\Big|_{T=T_n} \, ,
\label{eq:alpha}
\end{equation}
with $T_n$ being the nucleation temperature~\cite{Caprini:2019egz} whereas $\rho_{\text{rad}} \approx 35.12T^4$ is
the radiation energy density.
For this purpose, we perform LO
DR
of the SMEFT~\cite{Chala:2025aiz}, taking $g' =0$ and assuming the presence of the $\bar{c}_{\phi^6}, \bar{c}_{\phi^4D^2}, \bar{c}_{\phi^8}$ couplings in the UV at $\Lambda = 1\, \text{TeV}$ with sizes according to \cref{eq:all-scalings-soft}. We obtain $T_n$ using \texttt{FindBounce}~\cite{Guada:2020xnz}.
As outlined above, the inclusion of the NLO running of the effective parameters in the 3D EFT 
changes 
the PT
strength as much as
LO running (around 5\%--10\% for the example in scope).
To illustrate dependence on the specific values of the BSM couplings, we show the case if we take $\bar{c}_{\phi^4D^2}^{(1)}$ to be negative as a dashed line in \cref{fig:alpha}. 
It should be noted that the precise value of the shift of the PT strength with the inclusion of the NLO running is highly sensitive to the precise values of the beyond-SM couplings.

\section{Conclusions}
\label{sec:conclusions}
In this work, we have computed the leading corrections to the RG running of Wilson coefficients up to energy dimension four in the 3D EFT that describes the static high-temperature limit of BSM theories with heavy new physics. 

The computation, which involves evaluating two-loop diagrams with up to eight external legs, is considerably simplified by the energy scaling of parameters in 3D, the absence of subdivergences, and the small number of divergent integrals in dimReg.

We remark that just including the renormalizable sextic scalar interaction---which has not been addressed before---triggers running of the quartic and sextic operators at the leading, two-loop order. 

To briefly explore phenomenological implications, we focus on two cases of Higgs-field nucleation: one occurring at the soft scale, where the temporal vector modes are dynamical, and one below the soft scale, where the temporal vector modes are integrated out.
In the first case, a reference example of which is the tree-level nucleation, where the enhanced sextic Higgs coupling triggers the PT, we have found that the NLO beta function for $m_3^2$ typically modifies the LO result at the percent level, though the correction can become arbitrarily large near $\lambda_3\approx -0.375 g_3^2$, where the LO running vanishes. The impact on the scalar potential is even more pronounced (essentially, because the running of four-, six- and eight-Higgs scalar operators is parametrically of the same order as the LO running of $m_3^2$). For instance, while the relative shift of the broken minimum due to super-renormalizable running is at the percent level, including the NLO running increases this by more than an order of magnitude.
The shift on the energy difference between the broken and unbroken minima can also be modified by up to tens of percent, compared to the LO running, resulting in a change in the strength of the PT.

In the second case---nucleation below the soft scale, for example via a radiatively generated barrier---we find that the beyond-super-renormalizable running is subleading, 
reflected in smaller modifications of the potential.

The counterterms for the effective operators presented in this paper (e.g., \cref{eq:counterterms}) are also of direct use for future lattice simulations of the 3D EFT. In particular, they are essential for relating lattice and continuum parameters~\cite{Laine:1995np}. 
Beyond our $\overline{\text{MS}}$ running, divergences in the lattice regularization scheme need to be studied when extending the 3D EFT. In the super-renormalizable case present in the literature~\cite{Farakos:1994xh,Laine:1995np}, only the mass counterterm is linearly divergent. However, we find, for instance, that considering the renormalizable interaction $c_{\phi^6}$ results in a one-loop linearly divergent counterterm for the Higgs quartic (absent in dimReg), 
$\delta \lambda^L_3 = -12\Sigma c_{\phi^6}$, where $\Sigma = 0.252731/a$ with $a$ the lattice spacing. When accounting for higher-dimensional operators, the situation becomes more complex and as such, the full analysis of these divergences on the lattice is left for future work.

Altogether, our results enable more accurate determinations of cosmological PT parameters---and consequently of the resulting GW spectra---both within perturbation theory and in lattice simulations. In certain regions of parameter space, these corrections become quantitatively indispensable.

Finally, it is also worth commenting that we ignored gluon operators in the 3D EFT because they cannot arise from tree-level matching in DR,
though they could add small corrections to our RG results. Our results could also be extended with the inclusion of BSM particles within the EFT. In this respect, it could be very beneficial to compute the RGEs in the most general 3D EFT, in line with recent results in 4D~\cite{Guedes:2025sax}.

\section*{Acknowledgments} 
M.C. is supported by the European Research Council under Grant Agreement n. 101230200. A.D. is supported by the Spanish Research Agency (MICIU/AEI/10.13039/501100011033) under Grant No. CNS2024-154834. G.G. is partially funded by the Eric \& Wendy Schmidt Fund for Strategic Innovation through the CERN Next Generation Triggers project under Grant Agreement No. SIF-2023-004.
This work has received further funding from MICIU/AEI/10.13039/ 501100011033 (Grants No. PID2022-139466NB-C21/C22 and No. PID2024-161668NB-100) as well as from Junta de Andaluc\'ia (Grants No. FQM 101 and No. P21-00199).

\appendix

\crefalias{section}{appendix}

\section{Radiatively generated barrier}
\label{app:rgb}
For sufficiently large gauge coupling within the dimensionally reduced theory, a barrier can be radiatively generated, which enables a strong first-order PT.
It occurs when parametrically,
\begin{equation}
    m_{3}^2 T \sim \lambda_{3}T^2 \sim g_{3}^3 T^{3/2}.
\end{equation}
Moreover, as the running beyond the super-renormalizable contribution is triggered by the effective operators, starting from the sextic $c_{\phi^6}$, we further assume a sextic term of the scalar potential is as large as the quadratic and quartic terms, which leads to:
\begin{equation}
\label{eq:couplings-scalingUS}
    \tilde{g}^2 \sim g^{1.5} \frac{\Lambda}{T}.
\end{equation}

In this scenario, the nucleation of the scalar Higgs field occurs below the soft scale~\cite{Ekstedt:2022zro,Camargo-Molina:2024sde}, and the dimensionally reduced theory describing the nucleation is of the form of \cref{eq:lagrangian3d}, but without dynamical temporal components of the gauge fields.
Following the same steps as in the case of a tree-level barrier, discussed in the main text of this paper, we determine the running of the effective couplings:
\begin{equation}
    \label{eq:betafunctionsUS}
        \begin{split}
            \beta_{m_3^2} &=  \overbrace{-\frac{51}{16} g_3^4 - 9 g_3^2 \lambda_3 + 12 \lambda_3^2}^{\beta_{m_3^2}^{\text{LO}}} \\
            \quad &+ m^2_3 \bigg[ c_{\phi^4D^2}^{(1)}  \left( - 14 g_3^2 + 32 \lambda_3\right)  \\
            &\quad+  c_{\phi^4D^2}^{(2)}  \left(3 g_3^2 -8 \lambda_3\right) - 20 c_{\phi^2 W^2}  g_3^2 \bigg] \,, \\
            \beta_{\lambda_3} &= -c_{\phi^2 W^2}\left(\frac{39}{2} g_3^4 + 76 g_3^2 \lambda_3\right)   \\
            &\quad + c_{\phi^4D^2}^{(1)}\left(128 c_{\phi^6} m_3^2 +3 g_3^4 - 52 g_3^2 \lambda_3 +256 \lambda_3^2\right) \\
            &\quad + c_{\phi^4D^2}^{(2)}\left(-32 c_{\phi^6} m_3^2 - \frac{17}{16} g_3^4 +9 g_3^2 \lambda_3 - 60 \lambda_3^2\right) \\ 
            & \quad +   c_{\phi^6} \left(-18 g_3^2 + 96  \lambda_3\right) \,, \\
            \beta_{c_{\phi^6}} &= c_{\phi^6} \bigg[-168 c_{\phi^2 W^2} g_3^2 +c_{\phi^4D^2}^{(1)} \left(-84  g_3^2 + 1888 \lambda_3 \right)\\
            &\quad +c_{\phi^4D^2}^{(2)} \left( 18  g_3^2 - 448  \lambda_3 \right) + 204 c_{\phi^6} \bigg]  \\
            &\quad - 30 c_{\phi^8} \left(g_3^2 - 8 \lambda_3\right)\,, \\
            \beta_{c_{\phi^8}} &=12 c_{\phi^6} \left (296 c_{\phi^4D^2}^{(1)} c_{\phi^6} - 71 c_{\phi^4D^2}^{(2)} c_{\phi^6}+88  c_{\phi^8}\right)\,,\\
            \beta_{c_{\phi^4D^2}^{(1)}} &= 0\,, \\
            \beta_{c_{\phi^4D^2}^{(2)}} &=0\,,  \\
            \beta_{c_{\phi^2 W^2}} &=0\,,  \\ 
            \beta_{g_{3}^2} &= -\frac{28}{3} c_{\phi^2 W^2} g_3^4 \,. 
        \end{split}
\end{equation}
The corresponding counterterms and beta functions listed here could also be found in the Supplemental Material attached.

We estimate the size of the running of effective couplings,
\begin{equation}
\label{eq:runningscaling_rgb}
    \begin{split}
        \frac{\partial \log m_3^2}{\partial \log \mu_{3D}} &\sim \overbrace{g^1}^{\text{LO}} + g^{3.5} \frac{T}{\Lambda}\, , \\
        \frac{\partial \log \lambda_3}{\partial \log \mu_{3D}} &\sim g^2,\\
        \frac{\partial \log c_{\phi^6}}{\partial \log \mu_{3D}} &\sim \frac{\partial \log c_{\phi^8}}{\partial \log \mu_{3D}} \sim g^3\,, \\
        \frac{\partial \log g_3^2}{\partial \log \mu_{3D}} &\sim g^{5.5} \frac{T}{\Lambda}\,.
    \end{split} 
\end{equation}
and notice that in the case of the radiatively generated barrier, the running of $\lambda_3$ and $c_{\phi^6}$, triggered by operators beyond the super-renormalizable ones, is parametrically suppressed with respect to the LO running of $m_3^2$, in contrast to the case of the tree-level barrier (see \cref{sec:scalarpotential}). 
The NLO running becomes even smaller, once $c_{\phi_6}$ is not assumed to be enhanced. 

\section{Inclusion of the $U(1)_Y$}
\label{app:g1}
For completeness, we also calculate the contributions to the two-loop counterterms considering the extension of \cref{eq:lagrangian3d} below the soft scale (i.e., without temporal vector modes) to include a $U(1)_Y$ gauge symmetry with charge $g_3'$. The following operators are added:
\begin{equation}
    \begin{split}
        \mathcal{L}& \supset \frac{1}{4} B_{rs} B_{rs} \,\\ 
        &+ c_{\phi^2 B^2}|\phi|^2 B_{rs}B_{rs} + c_{\phi^2 WB}(\phi^\dagger \sigma^I \phi) W^I_{rs}B_{rs}\,.     
    \end{split}
\end{equation}
The counterterms in \cref{eq:counterterms} receive the following additional contributions:
\begin{equation}
    \begin{split}
        I^{-1}_{2}\delta Z_\phi &= -\frac{2}{3} c_{\phi^2B^2} g_3'^2 -2c_{\phi^2WB} g_3'g_3 \\
        &\quad-\frac{5}{3} c^{(1)}_{\phi^4D^2} g_3'^2 -\frac{1}{6} c^{(2)}_{\phi^4D^2} g_3'^2\, ,\\ 
        I^{-1}_{2}\delta Z_W &= \frac{4}{3} c_{\phi^2 WB} g_3' g_3 -4 c_{\phi^2W^2} g_3'^2    \,,\\
        I^{-1}_{2}\delta Z_B &= 4c_{\phi^2WB} g_3' g_3 -\frac{4}{3} c_{\phi^2B^2} g_3'^2 - 12 c_{\phi^2B^2} g_3^2 \,,\\
        I^{-1}_{2}\delta m^2_3&=\frac{5}{16}g_3'^4 + \frac{9}{8} g_3'^2 g_3^2 -3 g_3'^2\lambda \\
        &\quad+ m^2_3\biggl[6 c_{\phi^2 B^2} g_3'^2+6c_{\phi^2WB} g_3' g_3\\
        &\quad +3g_3'^2  c^{(1)}_{\phi^4D^2}+\frac{3}{2} g_3'^2 c^{(2)}_{\phi^4D^2} \biggr] \,, \\
        I^{-1}_{2}\delta \lambda &=  
        c_{\phi^2B^2} \left( \frac 92 g_3'^4+ \frac{13}{2} g_3'^2 g_3^2 - 24g_3'^2\lambda  \right)\\
        &\quad +c_{\phi^2 W B } \left(6 g_3'^3 g_3 +4 g_3' g_3^3 - 24 g_3' g_3 \lambda\right) \\
        &\quad +c^{(1)}_{\phi^4 D^2}\left(g_3'^4+2g_3'^2 g_3^2-14 g_3'^2 \lambda\right) \\
        &\quad +\frac{13}{2} c_{\phi^2W^2} g_3'^2 g_3^2 + 6c_{\phi^6}g_3'^2\\
        &\quad +c^{(2)}_{\phi^4 D^2}\left(\frac{15}{6}g_3'^4+\frac{23}{8} g_3'^2 g_3^2-4g_3'^2\lambda\right)\,,
        \\
        I^{-1}_{2}\delta c_{\phi^6} &=  c_{\phi^6} \Bigl[ 54 g_3'^2 c_{\phi^2B^2} + 54 c_{\phi^2 W B} g_3' g_3 \\
        &\quad + 33 c^{(1)}_{\phi^4 D^2} g_3'^2 - \frac{27}{2}c^{(2)}_{\phi^4 D^2} g_3'^2 \Bigr]\\
        &\quad + 10 c_{\phi^8}g_3'^2\,,
    \end{split}
\end{equation}
in which we only list those that receive nonzero contributions at the considered accuracy.

\bibliographystyle{apsrev4-2}
\bibliography{refs}

\end{document}